\documentclass{jfm}

\usepackage{graphicx}
\usepackage{newtxtext}
\usepackage{newtxmath}
\usepackage{natbib}
\usepackage{hyperref}
\hypersetup{
    colorlinks = true,
    urlcolor   = blue,
    citecolor  = green, 
}
\usepackage{pstricks}
\usepackage{pgf,tikz}
\usepackage{pgfplots}
\pgfplotsset{grid style={dashed,gray}}
\pgfplotsset{minor grid style={dashed,red}}
\pgfplotsset{major grid style={dotted,gray!80!black}}
\pgfplotsset{compat=1.14} 
\graphicspath{{Figures/}}
\usepackage{bm} 
\usepackage{color} 
\usepackage{caption,subcaption}
\newcommand{\sidecaption}[1]
{\raisebox{\abovecaptionskip}{\begin{subfigure}[t]{1.6em}
			\caption[singlelinecheck=off]{}
			\label{#1}
	\end{subfigure}}\ignorespaces}

\newcommand{\RomanNumeralCaps}[1]
\linenumbers

\title[Machine learning control]{Towards human-interpretable, automated learning of feedback control for the mixing layer}
\author[H.~Li, 
        G.~Y.~Cornejo Maceda,
       Y.~Li, 
        J.~Tan,
        M.~Morzy\'nski 
        \& B.~R.~Noack]
{Hao Li$^{1,2}$,
 Guy~Y.~Cornejo Maceda$^3$,
 Yiqing Li$^4$,
 Jianguo Tan$^1$$^\dagger$, \\
 Marek Morzy\'nski$^5$ \and\ 
 Bernd R. Noack$^{4,2}$%
\thanks{Email address for correspondence: jianguotan@nudt.edu.cn, bernd.noack@hit.edu.cn}}

\affiliation{
$^1$ Scicence and Technology on Scramjet Laboratory, \\ 
     National University of Defense Technology, 
     Changsha, Hunan Province, China
\\[\affilskip]
$^2$ Hermann-F\"ottinger-Institut, 
     Technische Universit\"at Berlin,\\
    M\"uller-Breslau-Stra{\ss}e 8, D-10623 Berlin, Germany
\\[\affilskip]
$^3$ LIMSI, CNRS, Universit\'e Paris-Saclay, B{\^a}t 507, rue du Belv\'ed\`ere, Campus Universitaire,
F-91403 Orsay, France \\[\affilskip]
$^4$ Center for Turbulence Control,
     Harbin Institute of Technology, Shenzhen, 
     Room 312, Building C, University Town, Xili, Shenzhen 518058, 
     China
\\[\affilskip]
$^5$ Chair of Virtual Engineering,
     Pozna\'n University of Technology,\\
     Jana Paw\l{}a II 24 street, 60-965 Pozna\'n, Poland
}

\pubyear{...}
\volume{...}
\pagerange{...}
\date{?; revised ?; accepted ?. - To be entered by editorial office}
\begin{document}
\maketitle

\begin{abstract}
We propose an automated analysis 
of the flow control behaviour
from an ensemble of control laws 
and associated time-resolved flow snapshots.
The input may be the rich data base of machine learning control (MLC)
optimizing a feedback law for a cost function in the plant.
The proposed methodology provides 
(1) insights into control landscape
which maps control laws to performance 
including extrema and ridge-lines,
(2) a catalogue of representative flow states 
and their contribution to cost function for investigated control laws
and 
(3) a visualization of the dynamics.
Key enablers are classification and feature extraction methods of machine learning.
The analysis is successfully applied to the stabilization of a mixing layer
with sensor-based feedback driving an upstream actuator.
The fluctuation energy is reduced by 26\%.
The control replaces unforced Kelvin-Helmholtz vortices with subsequent vortex pairing
by higher-frequency Kelvin-Helmholtz structures of lower energy.
These efforts target a human interpretable, fully automated analysis of MLC
identifying qualitatively  different actuation regimes, 
distilling corresponding coherent structures,
and developing a digital twin of the plant.
\end{abstract}

\section{Introduction}
\label{ToC:Introduction}
We augment an automated learning of flow control
by an analysis 
which may provide physical insights 
into qualitatively different regimes and the coherent structure dynamics.
The vast majority of active turbulence control studies
are performed in a model-free manner \citep{Brunton2015amr},
as control-oriented modeling of the actuation response 
from broadband frequency dynamics is still a challenge.

Steady or periodic operation of a single actuator
may be optimized by gradient-based approaches for one or few parameters, like extremum seeking.
However, the optimization of multiple actuators,
feedback with multiple sensors,
or more complex control dynamics 
naturally gives rise to rich search space of control laws.
The underlying regression problem aims to optimize a cost function
for a multiple-input multiple-output control law.
Machine learning provides powerful regression solvers
for complex optimization problems \citep{Fukami2020tcfd}.
Evolutionary algorithms \citep{Koumoutsakos2001aiaaj},
genetic algorithms \citep{Benard2016ef}, 
genetic programming \citep{Li2018am}, 
cluster-based control \citep{Nair2019jfm},
and reinforcement learning \citep{Rabault2019jfm}
may serve as examples.
Most optimizations require dozens to thousands of 
statistically representative  performance tests of control laws.
Typically, the best control law and few associated flow fields are present
while the vast majority of the created data base is ignored.
This study aims to use this data base 
for physically interpreting the learning process.
Starting point is the metric of attractor overlap \citep{Ishar2019jfm}.

This study focuses on sensor-based stabilization of the mixing layer
as notoriously difficult and well investigated control benchmark.
The mixing layer plays an important role 
for drag reduction, separation mitigation,
combustion enhancement, noise reduction, to name only a few configurations.
Open-loop investigations indicate that carefully calibrated multi-frequency actuation
may significantly enhance mixing increase \citep{Coats1997pecs}  
while high-frequency actuation stabilizes the mixing layer over a limited streamwise extent.
Curiously, a low-frequency actuation with about 63\% of the natural shedding frequency
is found to have a stabilizing effect on shear layers \citep{Pastoor2008jfm}.
The literature on in-time feedback control is sparse \citep{Wiltse2011ef}.
Phasor control may strongly destabilize the flow \citep{Parezanovic2016jfm}.
The direct stabilization may be achieved by opposition control,
i.e., when actuation and sensors are at the same location.
But it is notoriously difficult if sensors are several wavelengths downstream of the actuation,
as the mixing layer has a continuum of unstable frequencies.
The present study is based on a direct numerical simulation for
the two-dimensional mixing layer with upstream actuation 
and a downstream array of sensors.
The actuation leads to dynamically rich behaviour
while remaining computationally doable and physically interpretable.
The optimization is performed with linear genetic programming control \citep{Li2018am}.
The employed optimizer has cracked the most challenging control problems in the past
but this choice not essential for the proposed analysis methods.

In this paper, we aim to augment machine learning control
as automated and human-intepretable learning of feedback control
exploiting the complete data base of simulations. 
The paper is organized as follows. 
\S~ \ref{ToC:Plant} describes the numerical simulation and control of the mixing layer. 
\S~ \ref{ToC:Method}  describes the MLC system and the analysis methods. 
In \S~ \ref{ToC:Results}, the MLC for mixing destabilization and stabilization
is analyzed.
Finally, conclusions and outlook are provided in \S~\ref{ToC:Conclusions}.

\section{Mixing layer plant---Configuration and control problem}
\label{ToC:Plant}

In the present study,
an incompressible two-dimensional mixing layer 
with the velocity ratio $3:1$ is considered as the control plant. 
The flow is actuated by a small upstream volume force in the center of the mixing layer
and is monitored 
by a two-dimensional  array of $5 \times 5$ downstream velocity sensors.
The cost to be optimized is the cumulative fluctuation energy of these sensors.

The location vector is denoted by $\bm{x}=(x,y)$
where the $x$-axis is aligned with the streamwise direction 
and the $y$-axis denotes the transverse direction. 
The origin of the coordinate system is located 
at the left edge of actuation zone.  
The velocity field is represented by $\bm{u}=(u,v)$ 
where $u$ and $v$ are the streamwise and transverse component, 
respectively. 
All variables are assumed to be non-dimensionalized 
by the initial vorticity thickness  $\delta_{0}$, 
the low-speed velocity $U_{1}$ and density $\rho$. 
The convective velocity $U_{c}$ is approximated by the average
velocity of two sides $U_{c}=(U_{1}+U_{2})/2$.
The corresponding Reynolds number reads $Re=U_{c}\delta_{0}\rho / \mu$ 
where $\mu$ denotes the dynamic viscosity of the fluid. 
In this study,  $Re=200$ which defines a nearly inviscid dynamics.

\begin{figure}
	\centering
	\def\svgwidth{0.7\linewidth}
	\input{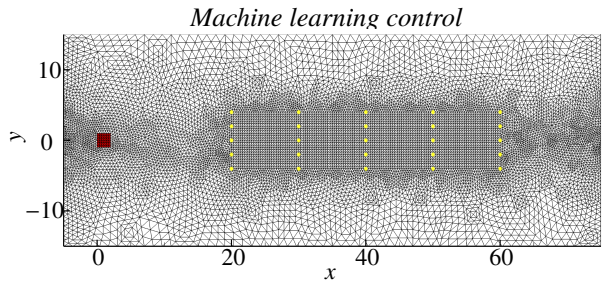}
	\caption{The computational domain. 
	$25$ sensors (yellow dots) are located at 
 	grid nodes to monitor the instantaneous velocity. 
	The transverse volume force as actuation is added in an upstream zone (red shade).}
	\label{Fig:SensorAndActuator}
\end{figure}
Figure \ref{Fig:SensorAndActuator} 
illustrates the rectangular computational domain $\Omega$, 
$x \in [-5,75]$, $y \in [-15,15]$
discretized  on a unstructured grid 
with $10237$ nodes. 
The inlet profile is a classical tanh profile
\begin{equation}
\label{Eqn:InletProfile}
u=2+\tanh \left( \frac{2y}{\delta_{0}} \right), 
\quad v=0, \quad \hbox{where} \quad \delta_{0}=1.
\end{equation}
The inlet velocity profile is perturbed
by a small stochastic excitation of $u$ component 
in the center at $y \in [-2,2]$ with 
a standard deviation of $0.01U_{c}$. 
This perturbation accelerates the evolution of the Kelvin-Helmholtz vortices.

A direct numerical simulation of the incompressible mixing layer 
is performed by an in-house solver based on
the finite element method. The numerical integration is second-order accurate in space
and fully implicit third-order accurate in time. 
The details can be referred to \citet{Ishar2019jfm}.  


The flow is manipulated 
by an upstream transverse unit volume force 
in the small domain $(x,y) \in  [0,2] \times [-1,1]$ 
(see the red shade in figure \ref{Fig:SensorAndActuator}). 
The volume force 
is driven by the actuation command $b$.
The absolute value of the command is limited by unity
to mimick experimental limits on actuation authority. 
The flow is monitored by an array of $N_s=25$ sensors 
for the streamwise velocity component.
The sensors (yellow dots in figure \ref{Fig:SensorAndActuator}) 
 are distributed in 
a rectangular area with 
$x  \in  \{ 20, 30, 40, 50, 60 \}$
and $y \in \{0,\pm 2, \pm 4 \}$. 
The computational mesh has been adjusted for an accurate placement of the sensors on the nodes. 
The feedback signal $\bm{s}(t)$ comprises all velocity components of these sensors.

The goal of sensor-based control is to minimize or maximize
the cumulative fluctuation energy of the sensors.
The signal fluctuation 
$s^{\prime}_i = s_i(t) - \langle s_i(t) \rangle_\tau$
is obtained based on the moving average 
$\langle s_i(t)\rangle_\tau = (1/\tau) \> \int_{t-\tau}^t s_i(t)$
over $[ t-\tau, t]$ with a time window $\tau=2/f$, 
where $f=0.1075$  is the dominant frequency of the natural mixing layer.
The averaged cumulative fluctuation energy reads
\begin{equation}
K = \sum_{i=1}^{N_{s}} \left\langle s_{i}^{\prime 2}(t) \right\rangle_{T} .     
\label{Eqn:K}
\end{equation}
Here, the averaging window is  $T=320$
corresponding to $8$ downwash times ($80/U_c=40$) through the whole compuational domain.       
For control purposes, 
the sensor-based control law shall 
maximize (minimize)  the kinetic energy $K$ ,
to destabilize (stabilize) the mixing layer. 
We define the cost function as 
$J_{d}=1/K$ ($J_{s}=K$) 
for destabilizing (stabilizing) the mixing layer. 

\section{Machine learning control augmented with data analysis}
\label{ToC:Method}

Machine learning control (MLC) \citep{Ren2020jh} 
has been applied to numerous experimental and numerical plants.
MLC can self-learn nonlinear multiple-input multiple-output feedback laws
minimizing a cost function using powerful methods of machine learning,
like reinforcement learning, 
genetic algorithms and genetic programming.
Here, we focus on linear genetic programming control (LGPC),
which has discovered the arguably most complex multiple-input multiple-output laws 
for distributed actuation of a turbulent jet \citep{Zhou2020jfm}.
Often hundred to thousand control laws are tested before convergence.
This data can provide valuable insight 
into the complexity of the control problem
and the associated flow physics.
In this section, we propose an analysis methodology for this data.
The goal is  to identify qualitatively different actuation mechanisms 
and their corresponding coherent structure dynamics.

Following \cite{Li2018am}, 
the control law is searched in the large space 
comprising sensor feedback and multi-frequency forcing and combinations thereof.
This  space includes 
many known stabilizing and destabilizing mechanisms,
like phasor control or low- and high-frequency actuation.
The ansatz for the control law reads
\begin{equation}
b = K(\boldsymbol{s}^{\prime}(t),\bm{h}(t)).
\label{Eqn:contrLaw}
\end{equation}
Here, $\bm{s}^\prime$ comprises $N_s$ sensor fluctuations introduced in  \S~\ref{ToC:Plant}.
And $\bm{h}$ includes harmonic signals
at natural frequency $f$, half and twice that frequency,
\begin{equation}
\label{Eqn:Harmonics}
\boldsymbol{h}(t)=\left[ \begin{array}{lllll}{\cos(2 \pi f t)} & {\cos( \pi ft)} & {\sin( \pi ft)} & {\cos(0.5 \pi ft)} & {\sin(0.5 \pi ft)}
\end{array}\right]^{\dagger}
\end{equation}
The harmonics include sines and cosines for the construction of phase differences.
These phase differences may have significant effect on mixing layers \citep{Coats1997pecs}.
Without loss of generality,  $\sin 2 \pi f t$ 
is omitted as only phase differences between different frequencies are of dynamic importance
and a phase in the first harmonics can be removed by a time shift.

The optimization of \eqref{Eqn:contrLaw} 
with respect to the cost $J$ is performed with LGPC
with typical parameters \citep{Li2018am} (see Appendix \ref{ToC:LGPC}).
The population size is $N_i=100$ and convergence is reached after $N_g$ generations.
LGPC leads to $N = N_i \times N_g$ control laws $K_i^n$ with cost $J_i^n$,
where $i$ and $n$ are the indices of the individual and generation, respectively.
The simulation of each control law leads to $N_t=800$ equidistantly sampled flow snapshots 
resolving the post-transient behaviour with time step $0.1$ representing two downwash times.

\begin{figure}
	\centering
        \includegraphics[trim=0 0 0 0,clip,width=0.8\linewidth]{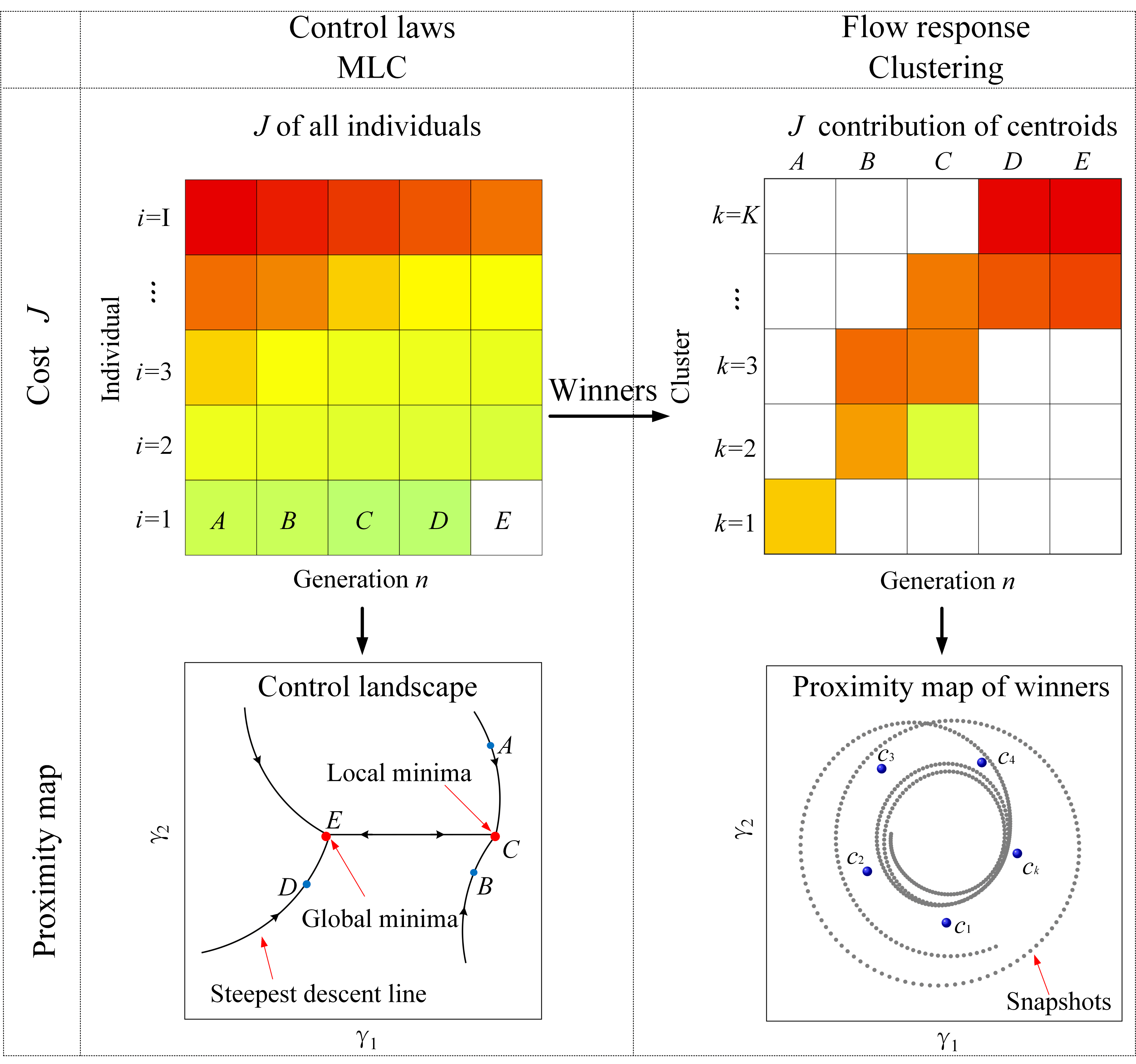}
	\caption{Principle sketch for augmented machine learning control. For details see text.}
	\label{Fig:Sketch}
\end{figure}
In the following, 
this data base is harvested for physical insights (see figure \ref{Fig:Sketch}).
Starting point is a completed MLC with $N_g$ generations of $N_i$ individuals (top left of figure \ref{Fig:Sketch}).
First, the tested ensemble of control laws 
are visualized in a control landscape
(bottom left of figure \ref{Fig:Sketch}).
In the following, 
$i$ denotes  the dummy index over all 
$N$ control laws $K_i$ and costs $J_i$.
Following \citet{Li2018am}, 
the distance between two control laws $K_i$ and $K_j$ 
is quantified by 
\begin{equation}
\label{Eqn:DistanceCL}
D_{ij} = M_{ij} + \alpha \left \vert J_i - J_j \right \vert 
\end{equation}
where $M_{ij} = \langle  \vert K_i - K_j \vert^2 \rangle_{i,j}$ 
is the time-averaged difference of the actuation commands
and the second term penalizes performance differences.
The penalization parameter $\alpha$ 
is chosen so that the maximum difference between control laws $M_{ij}$ 
equals the maximum difference $\alpha \vert J_i-J_j \vert$.
Thus, the symmetric configuration matrix $\bm{D}=(D_{ij})$ 
is based on control laws and cost functions.

The similarity and performance of the control laws
is visualized in a two-dimensional control landscape,
where each control law $K_i$ has an associated feature vector 
$\bm{\gamma}_i = \left[ \gamma_{i,1}, \gamma_{i,2} \right ]^\dagger$.
The distance between the feature vectors 
optimally approximates $D_{ij}$, i.e., 
$\Vert \bm{\gamma}_i - \bm{\gamma}_j \Vert^2 \approx D_{ij}^2 $.
Thus neighbouring (distant)  feature vectors represent 
similar (dissimilar) control laws or performances.
This goal is achieved with  
classical multidimensional scaling (CMDS) \citep{Cox2000book}.

As novel feature, 
the topology of the control landscape $J(\bm{\gamma})$
is illustrated with `steepest descent' lines 
which terminate in local or global minima.
Starting from $\bm{\gamma}_i$, 
the closest 15 neighbours are determined, corresponding to 2\% of the control laws.
Now, $\bm{\gamma}_i$ is connected  
to $\bm{\gamma}_j$  which has the lowest value of these neighbours.
The steepest descent continues from $\bm{\gamma}_j$  analogously
until the trajectory terminates in a minimum.
These steepest descent trajectories are determined for every feature vector.
The line width of a line from $i$ to $j$ 
increases with the number of passages from all trajectories.
Thus, deep valleys will be marked by thick lines.

The learning process of MLC with increasing generations
is illustrated by a comparison of the snapshots of the best individuals,
also called `winners' in the following.
The winner snapshots $\bm{u}^m$, $m=1,\ldots,M$ 
are coarse-grained into few representative centroids 
$\bm{c}_k$, $k=1,\ldots,K$ by a k-means++ algorithm (see Appendix  \ref{ToC:Method:Clustering}).
Each snapshot can be associated with its closest centroid.
Thus, snapshots are `binned' into clusters.
This coarse-graining allows to visualize and inspect all centroids 
and potentially give them a physical meaning
(top right of figure \ref{Fig:Sketch}).

The total fluctuation level is exactly given 
by 
$
J_i = \sum_{k=1}^K p_{i,k} J_{i,k},
$
where $p_{i,k}$ is the  population of a cluster $k$ by the $i$th control law,
and  $J_{i,k}$ corresponds to the fluctuation level associated with that cluster $k$.
Now, beneficial and less beneficial clusters are indicated by the local cost function $J_{i,k}$.
The machine learning process of a control law should 
increasingly populate increasingly better centroids 
while avoiding worse centroids, as observed by \cite{Nair2019jfm}.

The neighbourhood relation between snapshots and centroids
are visualized with another CMDS-based proximity map
(bottom right of figure \ref{Fig:Sketch}).
Thus, the temporal evolution of the Navier-Stokes simulatons 
can be mapped as trajectories 
in a two-dimensional plane with the centroids as `light houses'.
Cluster-based network modeling \citep{LiH2020jfm} provides an automated path to dynamic reduced-order models. 
The potential of this framework will become apparent in the result section
and will be critically discussed in the conclusions.

\section{Feedback control of the mixing layer}
\label{ToC:Results}
\begin{figure}
	\centering
	\sidecaption{subfig:a}
	\raisebox{-\height}{\def\svgwidth{0.41\linewidth}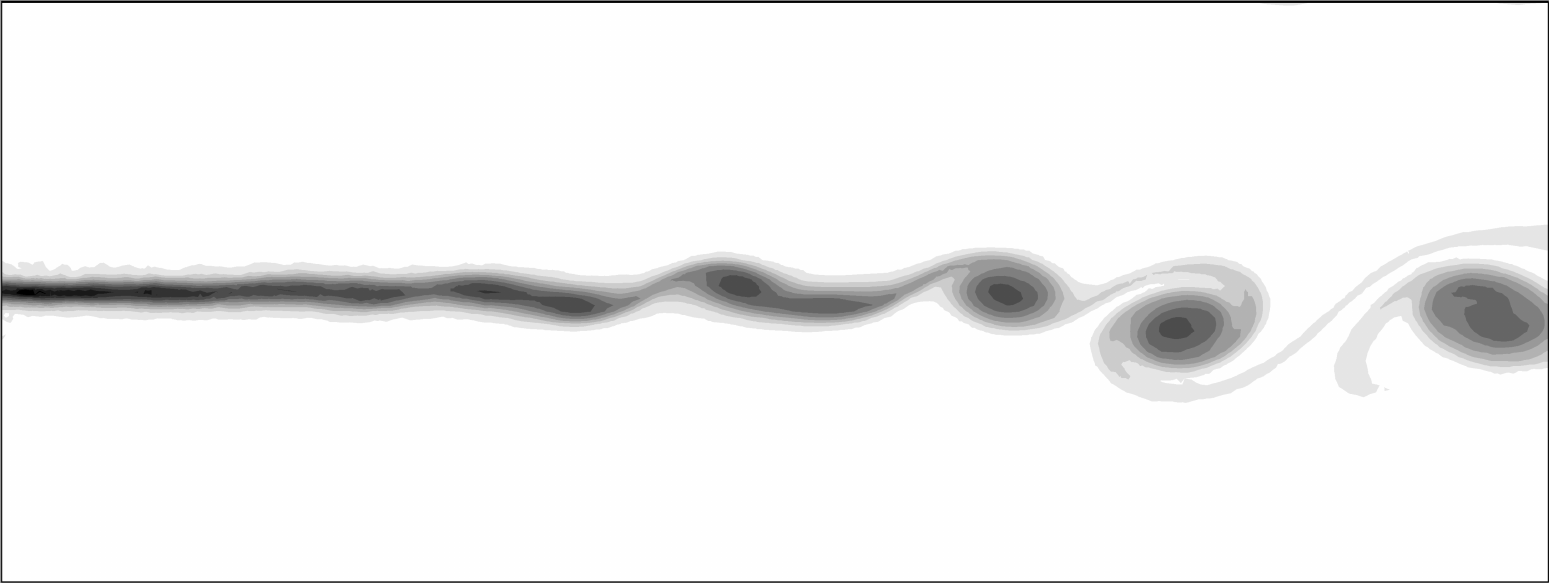}%
	\hfill
	\sidecaption{subfig:b}
	\raisebox{-\height}{\def\svgwidth{0.2\linewidth}
%
%
\begin{tikzpicture}

\begin{axis}[%
width=5.5cm,
height=2.0cm,
at={(0cm,2.5cm)},
scaled ticks=false, 
tick label style={/pgf/number format/fixed},
scale only axis,
bar shift auto,
xmin=0,
xmax=4.8,
yticklabels={\textcolor{white}{$Stab.$}, \textcolor{white}{$Nat.$}, \textcolor{white}{$Des.$}},
xlabel style={font=\color{white!15!black}},
xlabel={$J_{n}$},
ymin=0,
ymax=4,
ylabel style={font=\color{white!15!black}},
axis background/.style={fill=white},
]
\addplot[xbar, bar width=0.26, fill=white!20!black, draw=black, area legend] table[row sep=crcr] {%
0.74  1\\
1     2\\
2.5   3\\
};
\node[below right, align=left, draw=none]
(a) at (axis cs:2.6,3.5) {$Destabilization$};
\node[below right, align=left, draw=none]
(a) at (axis cs:1.1,2.5) {$Unforced \ Flow$};
\node[below right, align=left, draw=none]
(a) at (axis cs:0.8,1.5) {$Stabilization$};
\end{axis}

\end{tikzpicture}%
}
	\par
	\sidecaption{subfig:c}
	\raisebox{-\height}{\def\svgwidth{0.41\linewidth}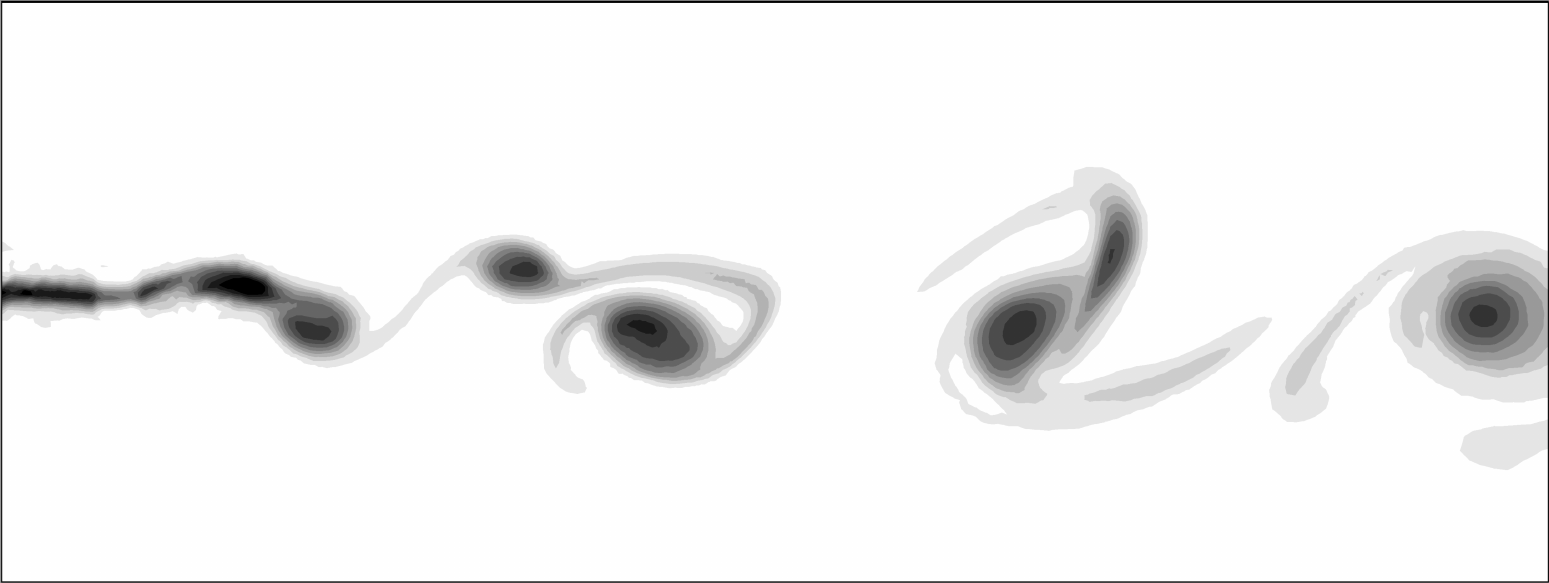}%
	\hfill
	\sidecaption{subfig:d}
	\raisebox{-\height}{\def\svgwidth{0.2\linewidth}\input{Figures/ActuationBestDestabilize.tex}}%
	\par
	\sidecaption{subfig:e}
	\raisebox{-\height}{\def\svgwidth{0.41\linewidth}
	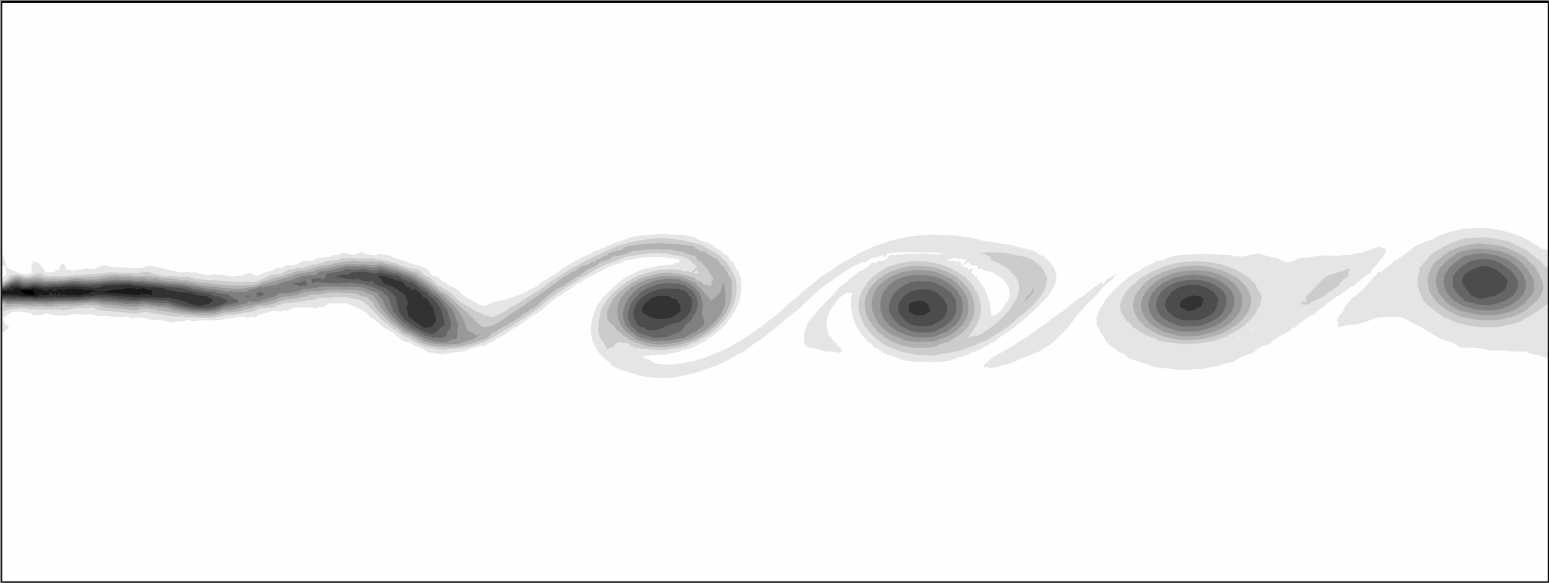}%
	\hfill
	\sidecaption{subfig:f}
	\raisebox{-\height}{\def\svgwidth{0.2\linewidth}\input{Figures/bStabilizeBestNEW.tex}}%
	\caption{Machine learning control results for destabilizing and stabilizing the mixing layer. 
	(\subref{subfig:a}) The spanwise vorticity of the natural regime.
	(\subref{subfig:b}) The performance of the linear genetic programming control.
	(\subref{subfig:c}) and (\subref{subfig:e}) respectively display the instantaneous spanwise vorticity of 
	destabilizing and stabilizing the mixing layer with the best control law displayed in  (\subref{subfig:d}) and (\subref{subfig:f}).}
	\label{Fig:ControlResults}
\end{figure}

First, linear genetic programming control (LGPC) is applied to stabilize and destabilize the mixing layer
with the ansatz \eqref{Eqn:contrLaw}.
A priori, the optimal strategy is far from being clear.
Open-loop low-frequency forcing may excite vortex pairing and thus destabilize the mixing layer.
Open-loop high-frequency forcing is reported to stabilize the flow.
Feedback mechanisms may or may not be better than periodic or multi-frequency forcing.
The mixing layer is notoriously difficult to stabilize 
with a continuum of unstable frequencies.

Figure \ref{Fig:ControlResults} presents the MLC results 
for minimizing and maximizing the fluctuation energy \eqref{Eqn:K}
after 6 generations with 100 individuals. 
The unforced benchmark (subfigure \subref{subfig:a})
shows Kelvin-Helmholtz vortices which tend to pair at the end of the domain.
For destabilization, MLC achieves the increase of fluctuation energy by a factor 2.5
(see subfigure \subref{subfig:b})
with the excitation of early multiple vortex pairings
(see subfigure \subref{subfig:c}).
MLC converges to a sensor-based feedback control 
and ignores the harmonic frequency input,
\begin{equation}
    b = - e^{s^\prime_{4}} / s^\prime_{21}.
\label{Eqn:OptimalCL}
\end{equation}
From subfigure \subref{subfig:d}, 
the actuation command is nearly periodic 'bang-bang' type,
i.e., assumes the maximum amplitude permitted by the constraint $\vert b \vert \le 1$.
The frequency is about half of the natural Kelvin-Helmholtz value.
The exponential factor remains positive.
Hence,  effectively, the control only listens to the sign of $s^\prime_{21}$
at the bottom left corner $x=20$, $y=-4$ of the sensor probes.
The feedback destabilization is similar to an MLC experiment
with upstream jets and a downstream hot-wire rake \citep{Li2018am}.

Stabilization of the mixing layer is a much harder task.
LGPC achieves a  $26 \%$ reduction of the fluctuation level (see figure \ref{Fig:ControlResults}\subref{subfig:b}).
Intriguingly, the control mechanism is not a delay of vortex formation
but early excitation of pure higher-frequency Kelvin-Helmholtz vortices 
which neither grow nor not pair 
(see figure \ref{Fig:ControlResults}\subref{subfig:e}).
Again, the best control law is of pure sensor feedback type,
\begin{equation}
    b = 0.075182-s^\prime_{7} \times \cos( s^\prime_{22} ).
\end{equation}
The actuation command induces a net upward force and feeds back
$s_7^{\prime}$, a sensor signal at $x=40$ and $y=2$.
The cosine factor is a small modulation from $s^{\prime}_{22}$ 
at the same downstream location $x=40$ but  on the other side $y=-2$ of the mixing layer.

\begin{figure}
	\centering
	\input{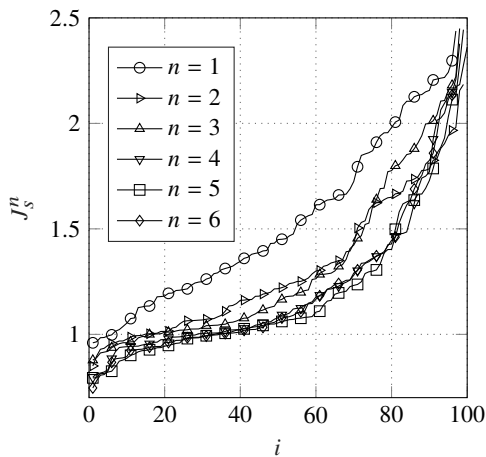}
	\caption{Mixing layer stabilization: Evolution of the normalized cost $J_{i}^{n}$ for the  
                 individuals $i=1,\ldots,100$ and generations $n=1,\ldots,6$.
		 For clarity, every fifth individual is displayed.}
		\label{fig:EvolutionLGPC}
\end{figure}
The learning curve of mixing layer stabilization (figure \ref{fig:EvolutionLGPC})
reveals the difficulty of the control task: 
Nearly all individuals of the initial generation $n=1$
are worse than  no forcing at all and few individuals 
of the subsequent generations beat the unforced benchmark of unity. 
The learning process is stopped at sixth generation.

\begin{figure}
	\centering
	\def\svgwidth{0.6\linewidth}
	\includegraphics[width=1\linewidth]{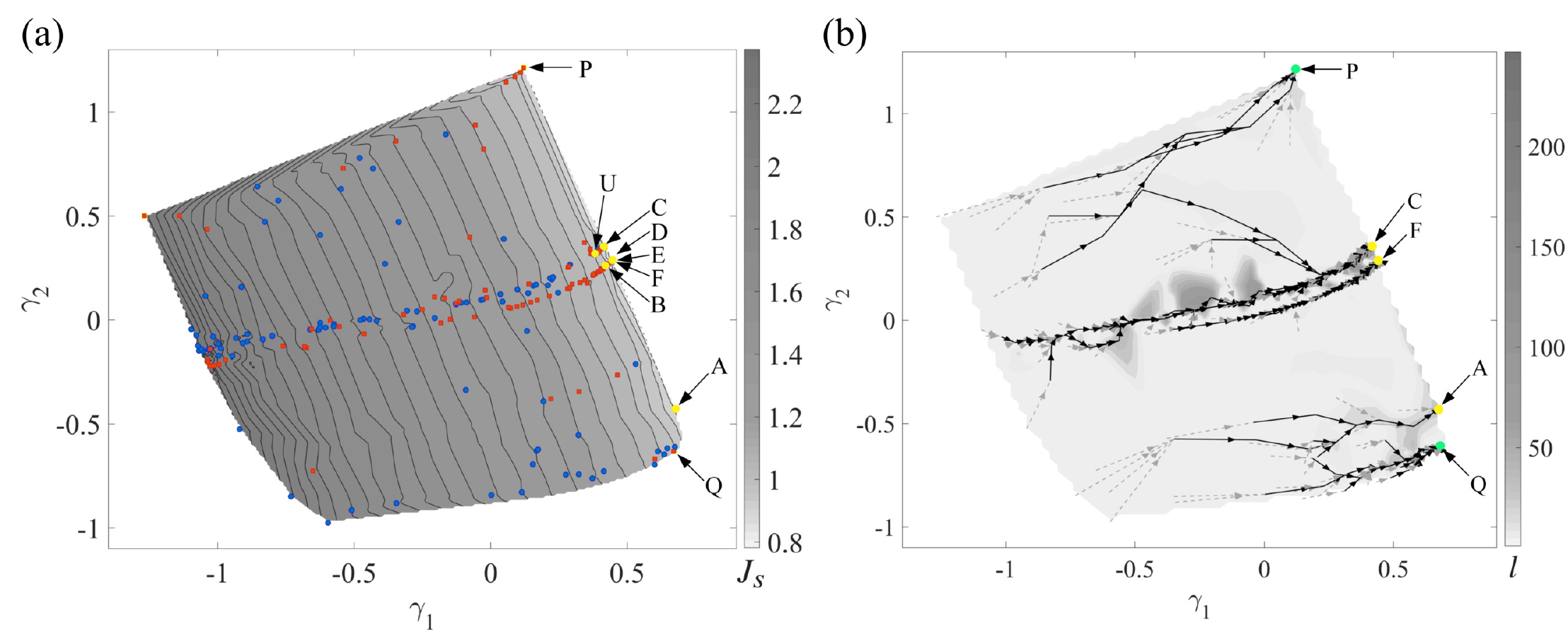}
	\caption{The evolution of machine learning control for stabilizing the mixing layer. 
	(a) Control landscape for $n=6$ generations. 
        (b) Ridgeline topology of the control landscape 
	via steepest descent lines. For details see text.}
	\label{Fig:ProxMap_Stabilize}
\end{figure}
In the following, we explore 
the data base generated by stabilizing LGPC with $ 100 \times 6 = 600$ simulations.
Figure  \ref{Fig:ProxMap_Stabilize}a shows the control landscape discussed in \S~\ref{ToC:Method}.
The  individuals of the first generation (blue dots)
span large portion of the landscape. 
The individuals in the sixth generation (red dots) 
assemble on the right side with increasing $\gamma_1$. 
The winners of the $n=1,\ldots, 6$ are denoted by yellow circles
marked $A$ to $F$, respectively.
The interpolated cost values are color-coded according to the legend.
Increasing $\gamma_1$ clearly reveals a better behavior of stabilizing mixing layer. 
The stabilization challenge may be appreciated by the closeness
of the best individuals $B$ to $F$ 
of generation  $n=2,\ldots, 6$  to the unforced flow marked by $U$.

In figure \ref{Fig:ProxMap_Stabilize}a, 
the control landscape is not populated evenly.
The individuals seem to populate lines as also observed in many other MLC studies.
The steepest descent lines (see figure \ref{Fig:ProxMap_Stabilize}b), discussed in \S~\ref{ToC:Method}, 
seem to offer an explanation.
 Line segments which are shared by at least $20$ of the search pathways are 
highlighted as black solid arrows and considered as the important pathway.
The control landscape has 5 local minima marked by red circles on the right side
associated with the winners $A$, $C$, $F$ and two suboptimal individuals $Q$ and $P$
from the last generation.
The steepest descent lines seem to quickly converge to 5 valleys leading to these 5 minima.
The valley may be a true ridgeline or a manifold which becomes one-dimensional after projection.
A closer investigation of higher feature coordinates $\gamma_i$, $i \ge 3$ may yield additional insight.
Summarizing, this analysis provides strong indication that stabilization is difficult,
was not achieved in the first Monte-Carlo generation and is associated with multiple minima.

\begin{figure}
	\centering
	\def\svgwidth{1\linewidth}
	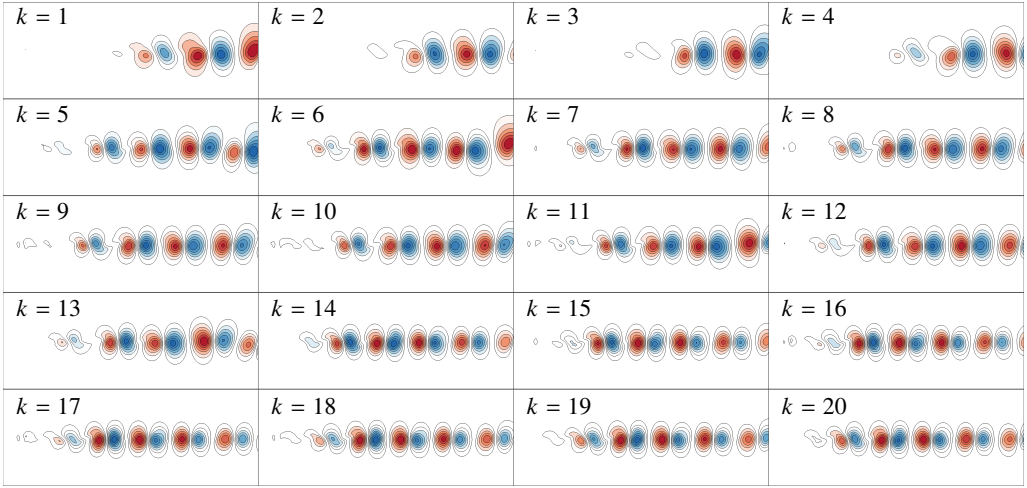
	\centering
	\caption{The $k=20$ centroids of the 
		natural flow and winners in each generation.}
	\label{fig:WinnerClusterCentroids}
\end{figure}


\begin{figure}
	\centering
	\def\svgwidth{0.6\linewidth}
	\includegraphics[width=1\linewidth]{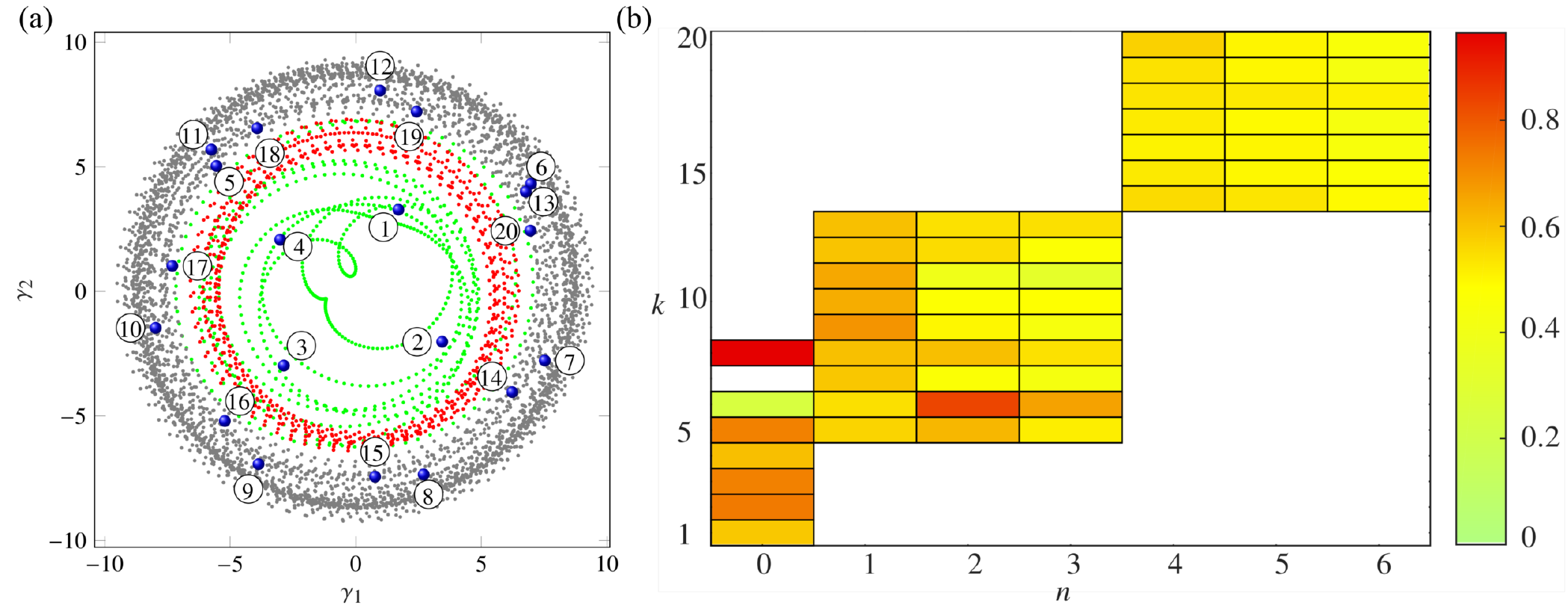}
	\caption{(a) Proximity map of the winners for stabilizing mixing layer simulations 
	showing the centroids (blue balls)
		and snapshots (dots). The green dots represent the snapshots of unforced attractor while the red 
		dots denote the snapshots from the attractor of best control law. Other gray dots represent 
		the winner snapshots of generation $2$ to $5$.
		(b) Fluctuation level of the clusters corresponding to the winners of each generation. 
		 $M$ denotes the index of generation.}
		 	\label{Fig:ClusterContribution}
\end{figure}

The control landscape may provide important insights
of dynamic domains with different behaviour.
In the following investigation, an understanding of the actuated coherent structure dynamics
shall be achieved from the unforced flow refereed to as $n=0$
and the winners of generations $n=1,\ldots,6$.
For every control law $800$ subsequent snapshots are investigated corresponding to two downwash times.
An exhaustive visualization and conclusive interpretation 
of all snapshots is typically beyond the scope of possibilities.
Here, we offer an automated coarse-graining into $K=20$ centroids,
These centroids are conveniently visualized 
in figure  \ref{fig:WinnerClusterCentroids}.
The similarity of the snapshots and centroids 
 is indicated in a 2D proximity map (figure \ref{Fig:ClusterContribution}a). 
The motion between centroids reveal the state transitions during the evolution
of machine learning control. 
The inner green dots denote the unforced flow populated cluster $k=1$ to $4$  
featuring long wavelength vortices 
and at cluster $k=5,6,7$ featuring short wavelength K-H vortices
while 8 describes the vortex shedding of the unforced flow. 
The winners of generation from $n=1$ to $3$ are described by the 
cluster from $k=5$ to $14$. 
The flow field of the winners are gradually tamed 
into a uniform limit cycle dynamics exhibiting 
K-H structures corresponding to the high-frequency forcing ($k=15$ to $20$).

Figure \ref{Fig:ClusterContribution}b sheds light
on the MLC optimization process 
and progressive migration 
from suboptimal to more optimal flow states.
The abscissa marks the considered winners $n=0,\ldots,6$
while colored boxes indicate the populated centroids $k=1,\ldots,20$.
The fluctuation contribution $K_k$ of each centroid is given by
\eqref{Eqn:Contribution}
\begin{equation}
K_{k}=\frac{\sum_{m=1}^{M} \chi^{m}_{k} \sum_{i=1}^{N_{s}} s_{i}^{\prime 2}(t) }{\sum_{m=1}^{M} \chi^{m}_{k}}.
\label{Eqn:Contribution}
\end{equation}
Here, $\chi_k^m$ is unity if the snapshot $m$ belongs to cluster $k$ and vanishes otherwise.
The sum of these centroidal cost values $K_k$ 
weighted by the population $p_k$ is exactly the total fluctuation level
$K = \sum_{k=1}^K p_k K_k$.
The color code of figure \ref{Fig:ClusterContribution}b indicates the $K_k$ value
of every centroid.
As expected, increasing performance is related to lower $K_k$ values.
Note that one centroid may have slightly different $K_k$ values 
for different control laws $n=0,\ldots,6$,
because the mean flow is different. 

\section{Conclusions}
\label{ToC:Conclusions}
We have augmented machine learning control (MLC)
by an automated analysis of the control law and associated flow data.
The analysis comprises 
(1) a control landscape 
with indications of local minima and ridgelines,
(2) a coarse-graining of snapshots from the best performing control laws 
into a small number of centroids, 
(3) an analysis of coherent structures represented by the centroids and the associated cost
and 
(4) a proximity map of the flow states visualizing the dynamics.
The augmented MLC has been successfully applied to the stabilization of the mixing layer.
The  control landscape indicated 
multiple local minima connected to rigdelines in long valleys.
The learning process of MLC reveals increasingly more efficient coherent structures.
A feedback law replaces unforced Kelvin-Helmholtz vortices with subsequent vortex pairing 
by  pure higher-frequency vortices which do not grow in streamwise direction. 
Thus, a reduction of the fluctuation energy by 26 \% was achieved.
In contrast, mixing layer destablization with MLC 
achieved an increase  of the fluctuation energy by a factor 2.5
by maximum amplitude feedback excitation of early multiple vortex pairing.

We see the proposed analysis as a start to a comprehensive
machine learning control and modeling strategy,
where self-learning of increasingly better control laws
provide the data base for
(1) an automated classification of qualitatively different dynamic regimes,
(2) an automated distillation of the most relevant coherent flow structures, and
(3) automated development of a control-oriented reduced-order model
approximating the full plant for a large range of control laws.
The control landscape and clustering are the starting point for tasks (1) and (2).
Preliminary results on cluster-based network models \citep{LiH2020jfm}
make them a likely candidate for task (3).
The automation of these tasks liberates time for 
deep first-principle based physics investigations.

\section*{Acknowledgements}

H.~L. wishes to acknowledge support from NSFC (No.91441121), the Graduate 
Student Research Innovation Project of Hunan Province (No.CX2018B027) and CSC (No. CSC201803170267). This work is supported by the ANR (grant 'FlowCon'), 
the DFG (grants SE 2504/2-1, SE 2504/3-1) 
and MNiSW (No.: 05/54/DSPB/6492).

\section*{Declaration of interests}

The authors report no conflict of interest.

\appendix 
\section{Linear genetic programming control}
\label{ToC:LGPC}
In the present study, the control goal for the mixing layer is to minimize (stabilization) or maximize (destabilization) 
the cumulative fluctuation energy $K$ of the sensors (see eq. \eqref{Eqn:K}). 
Following \citet{Li2017,Zhou2020jfm}, the control design is formulated as 
a regression problem: Find the best control law  
mapping the plant output to plant input which minimizes the cost function.
We have defined the cost function as $J_{d}=1/K$ ($J_{s}=K$) for destabilizing (stabilizing) the mixing layer.
The regression problem is solved by linear genetic programming (LGP) which can be referred to \citet{brameier2007linear}. 
\citet{duriez2017machine} successfully employed the genetic programming (GP) 
algorithm to mitigate flow separation and promote mixing.
LGP is a variant of GP and capable of finding simple, powerful and general regression method for
control laws of closed-loop control framework. 
We refer to this method as linear genetic programming control (LGPC).

\renewcommand{\arraystretch}{1.15}
\begin{table}
	\begin{center}
      \begin{tabular}{lc}
      Parameters & Values  \\ 
      Population size & $N_{i}=100$ \\
      	Tournament size & $N_{T}=7$ \\
      Elitism size    & $N_{e}=1$ \\
            	Replication probability & $P_{r}=10\% $ \\
      Crossover probability & $P_{c}=60\%$ \\
      Mutation probability &$P_{m}=30\%$ \\
      Operators        & $ +,-,x,\div,\sin,\cos,\tanh,\exp,\log$ \\
	Minimal instruction number & $5$ \\
	Maximal instruction number & $25$ \\
	Number of constants  & $N_{c}=6$ \\

	Constant range      & $[-1,1]$ \\
\end{tabular} 
      \caption{ LGPC parameters}
      \label{tab:LgpcParameters}
	\end{center}
\end{table}
The control law is formulated as a sequence of instructions operating on a set of sensors
$\bm{s}$ and harmonic functions $\bm{h}$, operators ($+$, $-$, $x$, $\div$, $\sin$,...) and constants. 
Table \ref{tab:LgpcParameters} lists the LGPC parameters for this study. 
LGPC initializes with the first generation containing $N_{i}$ individuals (i.e., control laws) randomly like in a Monte Carlo
method. Each control law will be evaluated in the numerical simulation plant and their performances on 
mixing layer destabilization and stabilization are quantified by the 
corresponding cost values $J$. The next generation is evolved by elitism and genetic operations.
The elitism is an operation which directly copy $N_{e}$ individuals of $n$ th 
generation with top ranking cost values to  the $(n+1)$th generation.
This ensures the new generations won't perform worse than the previous ones. 
The remaining $(N-N_{s})$ individuals of $(n+1)$th generation will be 
generated by three genetic operations including crossover, mutation and replication with specific selection
probabilities. The selected individual(s) for these three operations are the winner(s) of a 
tournament process. $N_{T}$ randomly selected individuals of $n$th
generation enter into a tournament process and the winner is determined by its cost value $J$. Crossover randomly exchanges 
substructures of two individuals, which is beneficial to breeding better potential individuals among well-performing
individuals. Mutation randomly changes a substructure in an individual. This operation might explore new local minima.
Replication copies an individual to next generation without any change. 
This results in a memory of the past generation.
These three genetic operators are stochastic 
in nature and have a fixed selection probabilities. We set the probabilities for three genetic 
operations $P_{r}=10\%$, $P_{c}=60\%$ and $P_{m}=30\%$ which are classical and result in good results in previous 
studies \citep{Li2017,Wu2018,Zhou2020jfm}. The procedure is iterated until some stopping criterion is reached, e.g., LGPC 
cannot find an individual with a much lower cost value $J$.    

To get a fast convergence for LGPC, a pre-selection of individuals is designed for all generations. Each 
individual is pre-evaluated based on the velocity fluctuation $\bm{s}^\prime$ from $N_{s}=25$ sensors of natural flow. 
An individual will be excluded if its actuation command $b$ is a constant during more than half time 
period, or a replicate of previous individuals. A new individual will be generated to 
replace the bad performer in the pre-evaluation step.

\section{Clustering as coarse-graining}
\label{ToC:Method:Clustering}

Cluster analysis lumps  similar objects into bins, called `clusters'.
This lumping of data is performed  in an unsupervised manner,
i.e., no advance labeling or grouping of the data has been performed.
We consider a sequence of velocity field snapshots denoted 
$\bm{u}^{m}\left(\bm{x} \right) :=\bm{u}\left(\bm{x}, t^{m}\right)$, 
$m=1,\ldots,M$ in a steady domain $\Omega$ from experiments or numerical simulations.
The velocity field is equidistantly sampled  with time step $\Delta t$,
i.e.\ the $m$th instant reads $t^{m}=m \Delta t$. These velocity snapshots are coarse-grained 
into $K$ clusters in terms of their similarities.  The representative states of each cluster is 
characterized by centroids $\bm{c}_{k}$.
The similarity of any two velocity fields $\boldsymbol{u}^{m}$, $\boldsymbol{u}^{n}$ 
is measured using Euclidean distance $D$, i.e.,

\begin{equation}
	D(\boldsymbol{u}^{m},\boldsymbol{u}^{n})=
	\sqrt{\int_{\Omega} \mathrm{d} x\|\boldsymbol{u}^{m}-\boldsymbol{u}^{n}\|^{2}}.
\label{Eqn:EuclideanDistance}
\end{equation}
The distance is calculated based on the norm associated with the Hilbert space
$\mathcal{L}^{2}(\Omega)$ of square-integrable functions. The whole computational 
domain $\Omega$ is employed for the integration.

A set of $K$ centroids $\boldsymbol{c}_{k}$ are randomly initialized and 
their performances are evaluated by the total inner-cluster variance $V$ of 
snapshots $\boldsymbol{u}^{m}$ with regard to the nearest
centroid $\boldsymbol{c}_{k}$,
\begin{equation}
V(\boldsymbol{c}_{1},\ldots,\boldsymbol{c}_{k})=\sum_{k=1}^{K} 
\sum_{\boldsymbol{u}^{m} \in \mathcal{C}_{k}} D\left(\boldsymbol{u}^{m}, \boldsymbol{c}_{k}\right).
\label{Eqn:Variance}
\end{equation}
Finally, a set of optimal centroids $\boldsymbol{c}^{opt}_{k}  (k=1, \ldots , K$) are determined to 
minimize the inner-cluster variance $V$ in  \eqref{Eqn:Variance}.
\begin{equation}
(\boldsymbol{c}^{opt}_{1}, \ldots, \boldsymbol{c}^{opt}_{K})=
\underset{\boldsymbol{c}_{1}, \ldots, \boldsymbol{c}_{K}}{\arg \min } 
V\left(\boldsymbol{c}_{1}, \ldots, \boldsymbol{c}_{K}\right)
\end{equation}
This optimization problem is solved using the iterative 
k-means algorithm \citep{MacQueen1967proc,Lloyd1982ieee}. 
The iteration will stop until the convergence is reached or while 
the variance $V$ is small enough. 
As no convergence to a global minimum can be guaranteed,
k-means++ performs 1000 k-means iterations 
 with different initial conditions and takes the best result.

The centroids characterize the typical flow patterns of each clusters, also called modes
in the ROM community.
The corresponding cluster-affiliation function maps 
a velocity field $\bm{u}$
to the index of the closest centroid,
\begin{equation}
k (\bm{u})  = \arg\min_i \Vert \bm{u} - \bm{c}_i \Vert_{\Omega},
\label{Eqn:ClusterAffiliation}
\end{equation}
This function defines cluster regions as Voronoi cells around the centroids
\begin{equation}
{\cal C}_i = \left \{  \bm{u} \in {\cal L}^2 ( \Omega )   \colon  k (\bm{u}) = i \right \}.
\label{Eqn:ClusterRegion}
\end{equation}
This function can also be employed to map a snapshot index $m$
to the representative cluster index 
$ k(m) := k\left ( \bm{u}^m \right ) $.
Alternatively, the characteristic function
\begin{equation}
\chi_i^m :=\left\{\begin{array}{ll}{1,} & \text{if} \quad i=k(m) \\ 
                                       {0,} & \text{otherwise}
                      \end{array}\right.
\label{Eqn:CharacteristicFunction}
\end{equation}
describes if the $m$th snapshot is affiliated with the $l$th centroid.
The latter two quantities are equivalent.

The number of snapshots $n_k$ in cluster $k$ is given by
\begin{equation}
n_k=\sum_{m=1}^{M} \chi_k^{m}
\end{equation}
The centroids are the mean velocity field of all snapshots in the corresponding cluster.
In other words, 
\begin{equation}
\bm{c}_k = \frac{1}{n_k} \sum_{\bm{u}^{m} \in \mathcal{C}_k} \bm{u}^{m}=\frac{1}{n_k} \sum_{m=1}^{M} \chi_k^{m} \bm{u}^{m}.
\label{Eqn:centroids}
\end{equation}
In the centroid visualizations (figure 6) , 
we accentuate  the vortical structures 
by displaying the fluctuations $\bm{c}_k - \overline{\bm{u}}$
around the snapshot mean $\overline{\bm{u}}$
and not the full velocity field $\bm{c}_k$.

In this study, we conduct the cluster analysis 
with  post-transient snapshots from the natural flow 
and the 6 winners of generations. 
From each of these 7 flows, also called attractors in the sequel, 
800 snapshots with sampling frequency of 10 are taken.
Note that $n=0, \ldots, 6$ is the index of the control law (attractor) 
and  $N_t=800$ is the number of the snapshots for each law.
All this 5600 snapshots are used for clustering.

Let $N_{i,k}$ be the number of snapshots 
of the $i$th attractor data in the $k$th data. Let $N_{i}=\sum_{k=1}^{K} N_{i,k}$
be the total number of snapshots of the $i$th attractor data. 
The probability that a snapshot belonging to the $i$th operating condition lies in the $k$ th cluster is estimated by
\begin{equation}
	p_{i,k}=\frac{N_{i,k}}{N_{i}}
\end{equation}
The probability describes the relative frequencies of cluster visits.

\bibliographystyle{jfm}

\end{document}